# Self-balancing of cell populations via martingale turnover with amplification

Tomoyuki Yamaguchi [*]
*Department of Basic Immunology, Research Institute, Nozaki Tokushukai Hospital, Osaka 5740074, Japan*

Adaptive control in biological systems, such as intestinal immunity, remains poorly understood despite detailed knowledge of underlying regulatory networks. We propose an alternative framework based on stochastic martingale turnover, in which cells proliferate through mutual competition and decay without cell-type-specific regulation. Through stochastic simulations and mathematical analysis, we show that this process autonomously generates balanced population compositions associated with low-decay probabilities. The compositional dynamics can be described as a random walk whose step lengths decrease in low-decay regions. Reduced decay leads to larger total population sizes and an increase in the number of compatible microscopic states, which in turn shapes the distribution of compositions under fluctuating conditions. More generally, the dynamics follow a modified Langevin equation, in which constant mass is replaced by a fitness-dependent effective mass proportional to the total population size. Thus, biological systems regulate resistance to change, not merely direction, in shaping their macroscopic behavior.

## I. INTRODUCTION

Living systems exhibit remarkable adaptability under fluctuating conditions. While molecular biology has identified many genetically encoded causal relationships among genes and cells, the principles underlying adaptive control—especially under biological noise—remain elusive [1–3]. A prominent example is intestinal immunity, where effector T cells ($T_{eff}$) protect against pathogens, while regulatory T cells ($T_{reg}$) suppress excessive inflammation [4]. Maintaining the balance between $T_{eff}$ and $T_{reg}$ is critical, yet the mechanisms achieving this regulation are not fully understood. Although ROR$\gamma$t-expressing antigen-presenting cells (ROR$\gamma$t$^+$ APCs) have been identified as key $T_{reg}$ activators [5–9], further processes must be invoked to explain how the system coordinates ROR$\gamma$t$^+$ and other APC subtypes that activate $T_{eff}$ [4,10]. The complexity of these regulatory networks suggests that exhaustive cataloging of molecular and cellular interactions may not suffice to explain their adaptive behaviors.

Inspired by advances in artificial intelligence, particularly reinforcement learning [11,12], we hypothesize that biological systems may implement analogous trial-and-error strategies for self-optimization [13–15]. In previous work, we proposed that stochastic epigenetic modifications—driven by amplification, noise, and decay—could underlie adaptive gene regulation [16]. There, gene expression autonomously stabilizes in favorable states where the epigenetic modification decays at low rates.

Here, we extend this framework to the autonomous balance of two cell populations. By formalizing the process as a stochastic differential equation (SDE), we show that biological adaptability naturally emerges in noise, relying on state-dependent resistance to change.

[*]Contact author: t.yamaguchi@tokushukai.jp

## II. METHODS

### A. Model

We investigate how two cell populations, such as $T_{eff}$ and $T_{reg}$ in intestinal immunity, can autonomously achieve balance in a stochastic setting. Extending our previous model of gene expression regulation [16], we describe the population dynamics of two cell types, I and II. Let $N_i(t) \in \mathbb{Z}_{\geqslant 0}$ denote the number of cells of type $i \in \{\text{I, II}\}$ at time $t$. Cell numbers evolve in discrete time through stochastic processes of increase and decrease (Fig. 1). We assume that the total population increases through two processes with constant event rates: proliferation under mutual competition (i.e., cell division) and extrinsic influx (e.g., migration). In the proliferation and decay processes, we simply assume that all cells—regardless of type—divide or die with equal probability at a given time, although these probabilities may vary temporally.

Proliferation is modeled as amplification under mutual competition. At each infinitesimal time step $\tau$, one cell divides with probability $a\tau$, with the dividing cell randomly selected from the total population (Fig. 1). Thus, the probability that a type-$i$ cell is selected to divide—causing $N_i$ to increase by one—is

$$P(N_i(t+\tau) = N_i(t) + 1 | \mathbb{N}(t)) = a\tau \frac{N_i(t)}{\sum_j N_j(t)}. \quad (1)$$

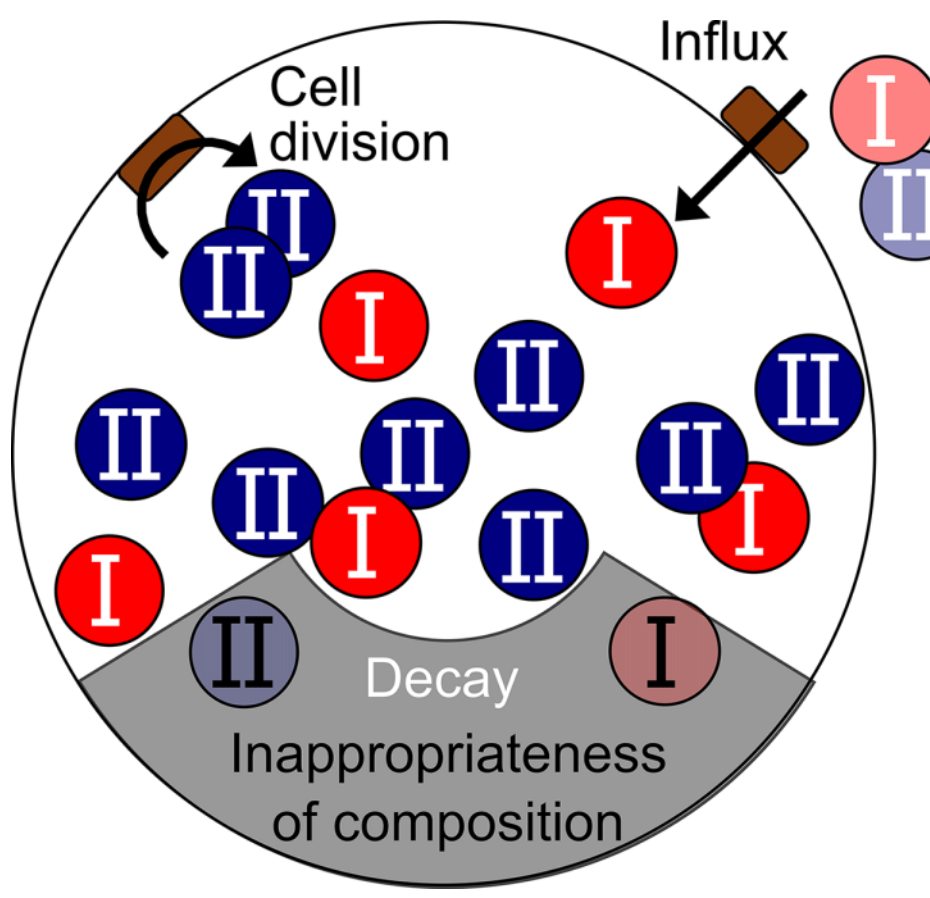

FIG. 1. Two-population model. Each cell stochastically proliferates, migrates, and dies in a semi-isolated region.

Here, $\mathbb{N}(t) \equiv (N_{\rm I}(t),\ N_{\rm II}(t))$. We consider biological situations in which growth triggers are provided with a constant probability and are immediately consumed by cells at a high reaction rate, as discussed in Appendix A.

This process resembles Pólya's urn: A randomly drawn ball (cell) is returned along with another of the same color. The resulting population ratio evolves as a martingale, a stochastic process whose expected future value equals its current value [17]. Notably, a bounded martingale converges almost surely to a random limit [18]. The distribution of this limit may be properly shaped in biological systems.

Influx, representing migration from the external environment, is modeled as a constant type-specific input. The probability that the number of type-$i$ cells increases by one in the interval $[t, t+\tau)$ is given by

$$P(N_i(t+\tau) = N_i(t) + 1|\mathbb{N}(t)) = b_i\tau, \tag{2}$$

where $b_i$ denotes the influx rate specific to cell type $i$.

Decay is modeled as a stochastic removal process. Each cell has an independent probability $\lambda(\mathbb{N}(t))\tau$ of being removed in the interval $[t, t+\tau)$. Consequently, the number of surviving type-$i$ cells follows a binomial distribution:

$$P(N_i(t+\tau) = n|\mathbb{N}(t)) = \binom{N_i(t)}{n}[1-\lambda(\mathbb{N}(t))\tau]^n\,[\lambda(\mathbb{N}(t))\tau]^{N_i(t)\,-\,n}, \tag{3}$$

where $n \in \{0, 1, \ldots, N_i(t)\}$. The decay rate $\lambda(\mathbb{N}(t))$ is assumed to increase with the deviation of the population ratio $X(t) \equiv N_{\rm I}(t)/\sum N_j(t)$ from $\alpha$. Thus, the ratio $\alpha$ corresponds to a biologically appropriate target composition that minimizes cell loss and promotes population persistence.

### B. Simulation methods

In the model describing the dynamics of cell numbers $N_{\rm I}(t)$ and $N_{\rm II}(t)$ under MTA (martingale turnover with amplification) and CSC (component-specific control) regimes, a series of three stochastic processes—proliferation [Eq. (1)], influx [Eq. (2)], and decrease [Eq. (3)]—was repeated $t = 10^5$ or $10^6$ time steps. Initial values were set to $N_{\rm I}(0) = N_{\rm II}(0) = 1$ in Fig. 2, unless otherwise specified [Sec. IIIA]. The target ratio for $N_{\rm I}(t)$:$N_{\rm II}(t)$ was set to $\alpha : (1-\alpha) = 1/3 : 2/3$.

In MTA regime, competitive proliferation occurs with probability $a = 0.1$ at each time step. When this intrinsic amplification proceeds, either cell type I or II is selected at the current ratio $N_{\rm I}(t)$:$N_{\rm II}(t)$, and the chosen type increases its count by one, only if $M(t) = N_{\rm I}(t) + N_{\rm II}(t) > 0$. Extrinsic increase via influx occurs with probability $b_{\rm add} = 0.001$. When this increase occurs, either cell type I or II is selected with a fixed probability $b_{\rm I}$:$b_{\rm II}$ = 1:1 to increase its number by one. In the decrease process, each cell is removed with probability $\lambda(\mathbb{N}(t))$ at each time step. The number $N_i(t+1)$ after decay is determined by a binomial distribution with $N_i(t)$ trials and success probability $1 - \lambda(\mathbb{N}(t))$. The same $\lambda(\mathbb{N}(t))$ is applied to both cell types. The decay probability $\lambda(\mathbb{N}(t))$ is a function of the difference between $X(t)$ and $\alpha$ with a minimum value of $10^{-4}$.

In mean squared error (MSE) model, $\lambda(\mathbb{N}(t)) = k\ {\rm MSE}_{X(t)} = 0.1[X(t)-\alpha]^2$ if $[X(t)-\alpha]^2 > 10^{-3}$; otherwise, $\lambda(\mathbb{N}(t)) = 10^{-4}$, where $X(t) \equiv N_{\rm I}(t)/\sum N_j(t)$.

In $\propto$ abs model, $\lambda(\mathbb{N}(t)) = k\ |X(t)-\alpha|$, with $k = 1/60$ if $|X(t)-\alpha|/60 > 10^{-4}$; otherwise, $\lambda(\mathbb{N}(t)) = 10^{-4}$. The parameter $k$ was adjusted to make the initial decay probability $\lambda(\mathbb{N}(0))$ same to that in the MSE model; $|1/2-1/3|/k = 0.1(1/2-1/3)^2 = 1/360$.

In WGM′ model, decay is based on the Kullback-Leibler divergence $D_{\rm KL}$ of the current composition $P = (X(t),\ 1 - X(t))$ from the target $Q = (\alpha,\ 1-\alpha)$. The survival probability is given by

$$\begin{aligned}&\exp(-kD_{\rm KL}(P||Q))\\ &\quad= \exp\left(-k\sum_j P_j \ln\left(\frac{P_j}{Q_j}\right)\right)\\ &\quad= \prod_j\left(\frac{Q_j}{P_j}\right)^{kP_j} = \left\{\left[\frac{\alpha}{X(t)}\right]^{X(t)}\left[\frac{1-\alpha}{1-X(t)}\right]^{1-X(t)}\right\}^k.\end{aligned} \tag{4}$$

WGM stands for the weighted geometric mean of the relative appropriateness across all cell types. Decay probability: $\lambda(\mathbb{N}(t)) = 1 - {\rm WGM}^k$ if $> 10^{-4}$; otherwise, $10^{-4}$. To match $\lambda(\mathbb{N}(0))$, solving $1-(2/3)^{k/2}(4/3)^{k/2} = 1/360$ gives $k = 2\ln(1-1/360)/\ln(8/9) \cong 0.0472$. See also Appendix B for its relevance to MSE and to biological systems.

In 4-level decay model, decay is stepwise $\lambda(\mathbb{N}(t))$ based on $0.1 \times$ MSE: $\lambda(\mathbb{N}(t)) = 10^{-4}$ if $[X(t)-\alpha]^2 \leqslant 10^{-3}$, $\lambda(\mathbb{N}(t)) = 0.0002$ if $10^{-3} < [X(t)-\alpha]^2 \leqslant 0.002$, $\lambda(\mathbb{N}(t)) = 0.002$ if $0.002 < [X(t)-\alpha]^2 \leqslant 0.02$, and $\lambda(\mathbb{N}(t)) = 0.02$ if $[X(t)-\alpha]^2 > 0.02$.

In CSC regime, cells increase only through influx with probability $a = 0$, $b_{\rm add} = 0.1$. When the influx occurs, cell type I or II is selected at a fixed ratio $b_{\rm I}$:$b_{\rm II}$ = 1/3:2/3 to increase its number by one. Both $N_{\rm I}(t)$ and $N_{\rm II}(t)$ decrease with a constant probability $\lambda(\mathbb{N}(t)) = 10^{-4}$ using a binomial distribution.

When comparing MSE simulation with theoretical prediction [Fig. 5(a) in Sec. III C], $\lambda(\mathbb{N}(t)) = 0.1[X(t)-\alpha]^2$ with a minimum value of $10^{-8}$. While $a = 0.1$, influx is disabled ($b_{add} = 0$). To prevent extinction, if $N_i(t) = 0$ after decay, it is reset to 1. Initial cell numbers are set to $N_{\mathrm{I}}(0) = N_{\mathrm{II}}(0) = aX(0)/\lambda(\mathbb{N}(0)) = 18$. Starting from $|X(0) - \alpha| = |1/2 - 1/3| = 1/6$, distances $|X(t) - \alpha|$ are recorded at time points $10^0$, $10^1$, $10^2$, …, $10^6$ across 1000 simulations.

## III. RESULTS

### A. Stochastic simulation

To assess whether the two-population model can regulate the population ratio, we performed stochastic simulations combining the three processes described by Eqs. (1)– (3) over $10^5$ repetitions. Under conditions favoring intrinsic proliferation ($a \gg b_i$; specifically, $a = 0.1$ and $b_i = 0.0005$) [Fig. 2(a)], the population ratio $[X(t),\ 1 - X(t)]$ autonomously approached the target $(\alpha,\ 1 - \alpha)$, when the decay probability was set to the maximum of $10^{-4}$ and $k$ times MSE, which is $k[X(t) - \alpha]^2$, with $k = 0.1$, $\alpha = 1/3$ [Fig. 2(b)]. Notably, the competition/$\sum N_j(t)$ in Eq. (1), along with the small but nonzero influx $b_i$ in Eq. (2), prevents divergence, extinction, or dominance by a single-cell type, as shown in Appendix A and Ref. [16]. We further tested alternative decay functions, including the absolute difference [Fig. 2(c)], the complement of the exponential Kullback-Leibler divergence, which is equivalent to a WGM of the ratio of target over present composition [Fig. 2(d) and Appendix B] and a four-level discretized decay probability based on MSE [Figs. 2(e) and 2(f)]. All forms effectively stabilized the population ratio near the target across most time points [Figs. 2(b)–2(f)] and across replicates at $t = 10^5$ [Fig. 2(g)].

The framework also encompasses a deterministic control regime, in which cell numbers increase exclusively through influx, not proliferation [Figs. 2(h) and 2(i)]. For example, with $a = 0$, $\lambda(X(t)) = 10^{-4}$, $b_{\mathrm{I}} = 0.033$, and $b_{\mathrm{II}} = 0.067$, the ratio $N_{\mathrm{I}}(t){:}N_{\mathrm{II}}(t)$ directly approached the target 1:2 without initial lag or transient fluctuations [Fig. 2(h)], and exhibited minimal variance at $t = 10^5$ [CSC in Fig. 2(g)]. However, this regime lacks adaptability, as it requires prior specification of the correct influx ratio $b_{\mathrm{I}}{:}b_{\mathrm{II}}$.

Taken together, these results demonstrate that two distinct mechanisms can achieve an appropriate population ratio, denoted here by $N_{\mathrm{I}}(t){:}N_{\mathrm{II}}(t) \cong \gamma{:}(1-\gamma)$, along with the corresponding increase ratio $I_{\mathrm{I}}{:}\ I_{\mathrm{II}} \cong \gamma{:}(1-\gamma)$. One is CSC, in which the increase rates are fixed as $I_i = b_i$. CSC requires a proper influx ratio $b_{\mathrm{I}}{:}b_{\mathrm{II}} \cong \gamma{:}(1-\gamma)$. The other is MTA, in which increase rates are endogenously determined as $I_{\mathrm{I}}{:}\ I_{\mathrm{II}} = N_{\mathrm{I}}(t){:}N_{\mathrm{II}}(t)$. MTA requires that the decay rate of each component decreases under appropriate compositions by setting $\alpha \cong \gamma$. Under this condition, the composition $N_{\rm I}(t){:}N_{\rm II}(t) \cong \gamma{:}(1-\gamma)$ becomes self-maintaining, as detailed in Secs. III B and III C.

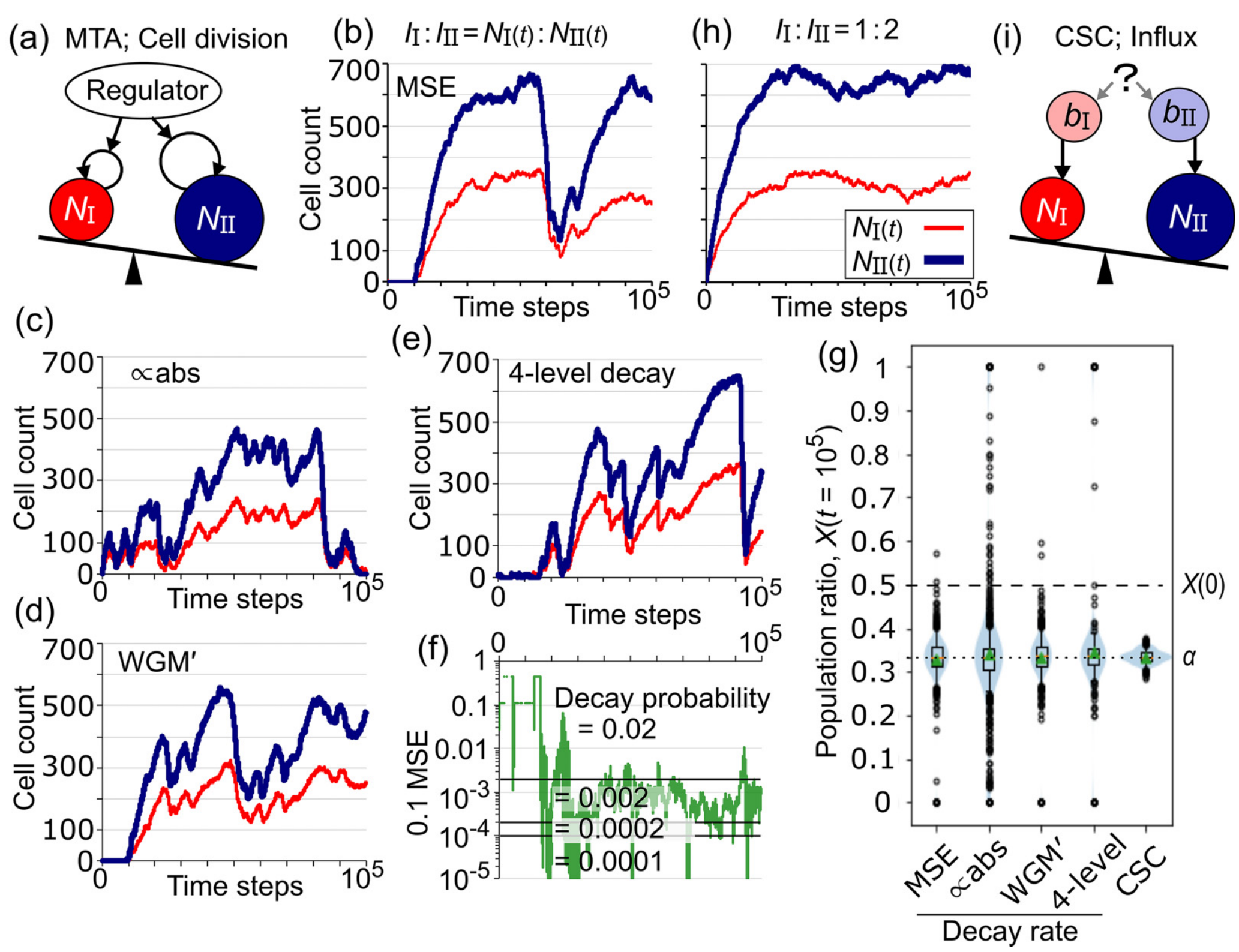

FIG. 2. Stochastic simulations of two-population model. (a)–(f) In MTA regime, cell numbers $N_{\mathrm{I}}(t)$ and $N_{\mathrm{II}}(t)$ increase mainly via competitive proliferation at the current $N_{\mathrm{I}}(t) : N_{\mathrm{II}}(t)$ ratio, and decrease at a rate determined by the fitness of the current composition (the target ratio $\alpha = 1/3$). (b) In "MSE," decay rate is $0.1[X(t) - 1/3]^2$. (c) In "∝ abs," decay rate is $|X(t) - 1/3|/60$. (d) In "WGM′," a weighted geometric mean is used for decay rate. (e), (f) In "4-level," four discrete probabilities based on MSE are used, as shown in panel (f). (g) Box and violin plots of $X(t)$ at $t = 10^5$ from 1000 simulations, excluding cases where $\sum N_j(t = 10^5) = 0$. The numbers of cases with $\sum N_j(t) \neq 0$ were 997 (MSE), 999 (∝ abs), 993 (WGM′), 968 (4-level), and 1000 (CSC). Green marks indicate the means. (h), (i) In CSC regime, cells increase via influx at a fixed 1:2 ratio. The mechanism controlling this influx is essential, but lies outside the model.

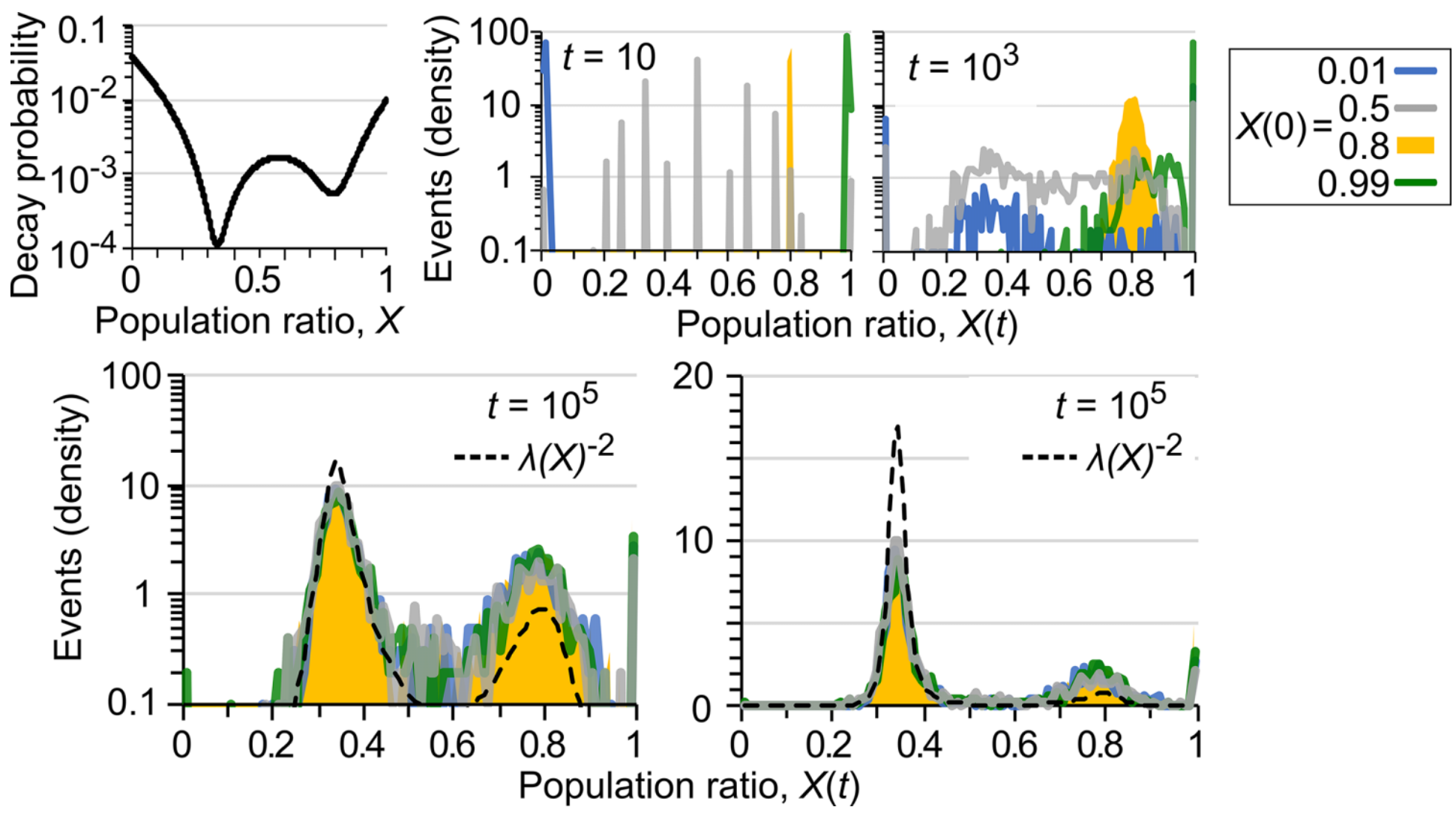

FIG. 3. Decay-dependent distribution pattern. In MTA regime ($a = 0.1$ and $b_i = 0.0005$), decay probabilities are set to be low around $X(t) = 1/3$ or 0.8; $\lambda(\mathbb{N}(t)) = 0.5[(X(t) - 1/3)^2 + 10^{-3}][(X(t) - 0.8)^2 + 5 \times 10^{-3}]$. Under four initial conditions: $(N_{\rm I}(0),\ N_{\rm II}(0)) = (1,\ 99)$, (1, 1), (160, 40), and (99, 1), 1000 simulations were performed. Histograms of $X(t)$ with bin width = 0.01 are shown after normalization. Dotted lines in the panels $t = 10^5$ show the theoretical stationary probability density in the drift-free limit: $1/[4.6 \times 10^6 \lambda(X)^2]$.

The adaptive behavior in MTA does not require cell-type-specific drift, tuned initial conditions, or knowledge of a target composition. When a more complex decay function—representing a biological fitness landscape with multiple low-decay regions—was applied in the MTA framework, population ratios across 1000 simulations converged to the same distribution at $t = 10^5$, regardless of initial conditions (Fig. 3). Notably, cells in this model simply experience a decay probability determined by the current state, without performing gradient descent on the landscape or having knowledge of other possible states or a global optimum. The distribution probability after repetitions reflects the decay rate and the resulting total population size at the population ratio $X$, as described in Sec. III C. Thus, the MTA framework provides a mechanism for autonomous population balancing under fluctuating biological conditions.

### B. Stochastic differential equation

We next mathematically formalize the regulatory dynamics of MTA and CSC (see also Appendix D). By taking an infinitesimal time interval $\tau$, we assume that the probability of two or more cells increasing or decreasing within $\tau$ is negligible compared to the probability of a single event [19]. From Eqs. (1)–(3), the expected change (denoted by $\langle\cdot\rangle$) in the number of type-$i$ cells over $\tau$ is

$$\frac{\langle N_i(t+\tau) - N_i(t)\rangle}{\tau} = a\frac{N_i(t)}{\sum_j N_j(t)} - \lambda(\mathbb{N}(t))N_i(t) + b_i. \tag{5}$$

This equation matches prior models of gene expression [16] and immune regulation [20].

Because the amplification parameter $a$ and decay rate $\lambda(\mathbb{N}(t))$ are common to both cell types at a given time point, summing Eq. (5) over all $i$ gives the expected change in the total cell number $M(t) \equiv \sum N_j(t)$:

$$\frac{\langle M(t+\tau) - M(t)\rangle}{\tau} = a - \lambda(X(t))M(t) + b, \tag{6}$$

where $b \equiv \sum b_j$, and the per-cell decay rate is expressed as $\lambda(X(t))$ by assuming that the decay probability of each cell depends on the composition, not on the total population size $M(t)$. Notably, Eq. (6) resembles the general form of stochastic degradation turnover in biological systems [21]. Beyond this, our model explicitly separates intrinsic proliferation [the first term of Eq. (5)] from extrinsic influx (the third term) in the increase process of stochastic turnover.

Given that $0 < \lambda(X(t))\tau < 1$, Eq. (6) shows that the expected total population size converges to

$$\langle M(t)\rangle \cong \frac{a+b}{\lambda(X(t))}. \tag{7}$$

Because many independent decay events occur at each step, $M(t)$ relaxes much faster than the composition $X(t)$. Hence, $M(t)$ can be approximated by the quasisteady value given by Eq. (7).

Since the key variable under regulation is the composition $X(t)$, we now examine its evolution over an infinitesimal interval $\tau$. When a cell increases, $X(t)$ increases by $[1 - X(t)]/[M(t)+1]$ with probability $[aX(t) + b_{\rm I}]\tau$, and decreases by $X(t)/[M(t)+1]$ with probability $\{a[1 - X(t)] + b_{\rm II}\}\tau$. When a cell decreases, $X(t)$ increases by $X(t)/[M(t) - 1]$ with probability $\lambda(X(t))M(t)[1 - X(t)]\tau$, and decreases by $[1 - X(t)]/[M(t) - 1]$ with probability $\lambda(X(t))M(t)X(t)\tau$.

These events yield the following expectation and variance $V(\cdot)$:

$$\frac{\langle X(t+\tau)-X(t)\rangle}{\tau}=\frac{b}{M(t)+1}[\beta-X(t)], \tag{8}$$

$$\frac{V(X(t+\tau)-X(t))}{\tau}=\frac{aX(t)[1-X(t)]}{[M(t)+1]^2}+\frac{\lambda(X(t))M(t)X(t)[1-X(t)]}{[M(t)-1]^2}+\frac{b[X^2(t)-2\beta X(t)+\beta]}{[M(t)+1]^2}, \tag{9}$$

where $\beta\equiv b_{\rm I}/b$. Equation (8) shows that MTA with $b=0$ satisfies the martingale condition $\langle X(t+\tau)\rangle=X(t)$.

Assuming $M(t)\gg 1$ and $\lambda(X(t))M(t)\cong a+b$, the variance Eq. (9) simplifies to

$$\frac{V(X(t+\tau)-X(t))}{\tau}\cong\frac{2aX(t)[1-X(t)]+b[X(t)-2\beta X(t)+\beta]}{M(t)^2}. \tag{10}$$

The variance near the target composition $X(t)\cong\alpha$ largely depends on $M(t)$, particularly when $\lambda(X(t))$ becomes sufficiently small around this composition. Under these conditions, the variance is further approximated by using a constant parameter $\sigma$:

$$\frac{V(X(t+\tau)-X(t))}{\tau}\cong\frac{\sigma^2}{M(t)^2}, \tag{11}$$

where

$$\sigma^2\equiv 2a\alpha(1-\alpha)+b(\alpha+\beta-2\alpha\beta). \tag{12}$$

These approximations hold when both populations are sufficiently large and the composition remains near the target, i.e., $M(t)\gg 1$, $X(t)\cong\alpha$, and $\lambda(X(t))M(t)\cong a+b$. They do not apply in situations such as initial lag phases or catastrophic collapse.

Under these conditions, the composition dynamics follow an SDE

$$\frac{dX(t)}{dt}=\frac{b}{M(t)}[\beta-X(t)]+\frac{\sigma\,\eta(t)}{M(t)}, \tag{13}$$

where $\eta(t)$ is temporally uncorrelated Gaussian white noise. Substituting Eq. (7) for $M(t)$ gives

$$\frac{dX(t)}{dt}=\frac{b}{a+b}\lambda(X(t))\left[\beta-X(t)+\frac{\sigma}{b}\eta(t)\right]. \tag{14}$$

This formulation clarifies how the decay rate $\lambda(X(t))$, intrinsic proliferation $a$, and extrinsic influx $b$ shape composition dynamics under stochastic fluctuations (Fig. 4 and Appendix C). In CSC, where $b\gg a$, the noise term $\sigma\eta(t)/b$ becomes negligible, stabilizing $X(t)$ near $\beta$. In contrast, MTA, where $a\gg b\cong 0$, relies on noise-driven fluctuations to allow $X(t)$ to settle near the point where $\lambda(X(t))\cong 0$, i.e., $X(t)\cong\alpha$.

Multiplying both sides of Eq. (13) by $M(t)dt$ gives a Langevin-like form:

$$M(t)\,dX(t)=b[\beta-X(t)]dt+\sigma\,dW(t), \tag{15}$$

where $W(t)$ is a Wiener process (Brownian motion) satisfying $\langle W(t)\rangle=0$ and $\langle W(t)^2\rangle=t$.

Importantly, Eq. (15) reduces to the conventional Langevin equation when the effective mass $M(t)$ is constant. In physical systems, mass is typically conserved when a particle moves under biased stochastic forces and kinetic friction. In contrast, in a thermodynamically open biological system with continuous turnover, the mass of total population $M(t)$ varies according to the current fitness. Thus, physical Langevin dynamics can be viewed as a special case of a more general stochastic formulation in which "resistance to change" is dynamically regulated. In other words, Eq. (15) highlights the importance of state-dependent resistance to change for adaptive behaviors in biological systems.

### C. Analytical solution

We elucidate how composition $X(t)$ distribution forms a particular pattern in the MTA framework, where drift force is almost absent. Substituting $b=0$ into Eqs. (14) and (15) gives

$$dX(t)=\frac{\sigma}{M(t)}W(t)\quad\text{or}\quad dX(t)=\frac{\sigma\lambda(X(t))}{a}W(t). \tag{16}$$

This shows that the behavior of $X(t)$ becomes a random walk in which the next step length varies depending on the present position. Because composition is naturally confined to a bounded interval $[0\leqslant X(t)\leqslant 1]$, we consider reflecting boundary conditions in the corresponding Fokker-Planck equation. In the MTA simulations, a small influx term (implemented as the avoidance of extinction) prevents trajectories from sticking to the boundaries. Then, the stationary density in the drift-free limit satisfies $p_\infty(X)\propto M(X)^2\propto\lambda(X)^{-2}$. This prediction is consistent with the distribution observed in simulations at $t=10^5$ in Fig. 3.

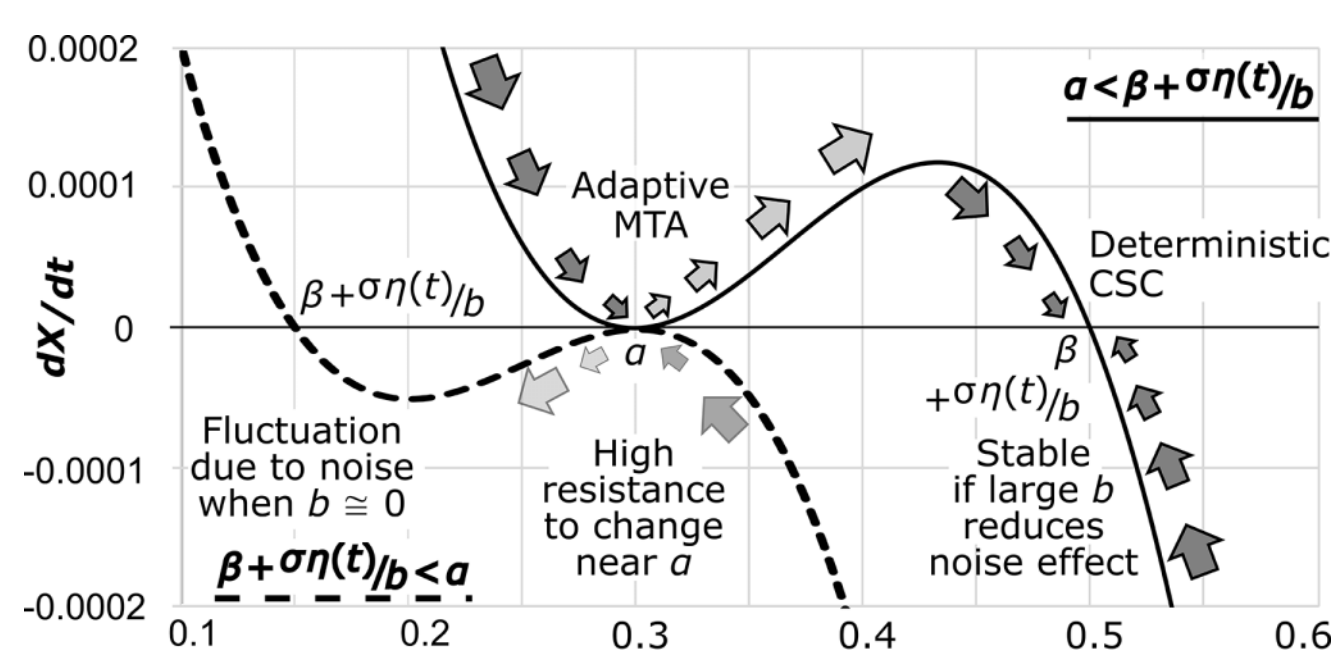

FIG. 4. Noise effects in MTA and CSC. The rate of change $dX(t)/dt$ is plotted against $X(t)$ under two conditions in Eq. (14); $\lambda(X(t))=[X(t)-\alpha]^2$, $b/(a+b)=0.1$, $\alpha=0.3$, and either $\beta+\sigma\eta(t)/b=0.5$ (solid curve) or 0.15 (dotted curve). The point $\beta+\sigma\eta(t)/b$ acts as a stable fixed point only when the noise term is sufficiently suppressed by large $b$. The saddlelike point $\alpha$ has both stable and unstable sides on each curve. See also Appendix C.

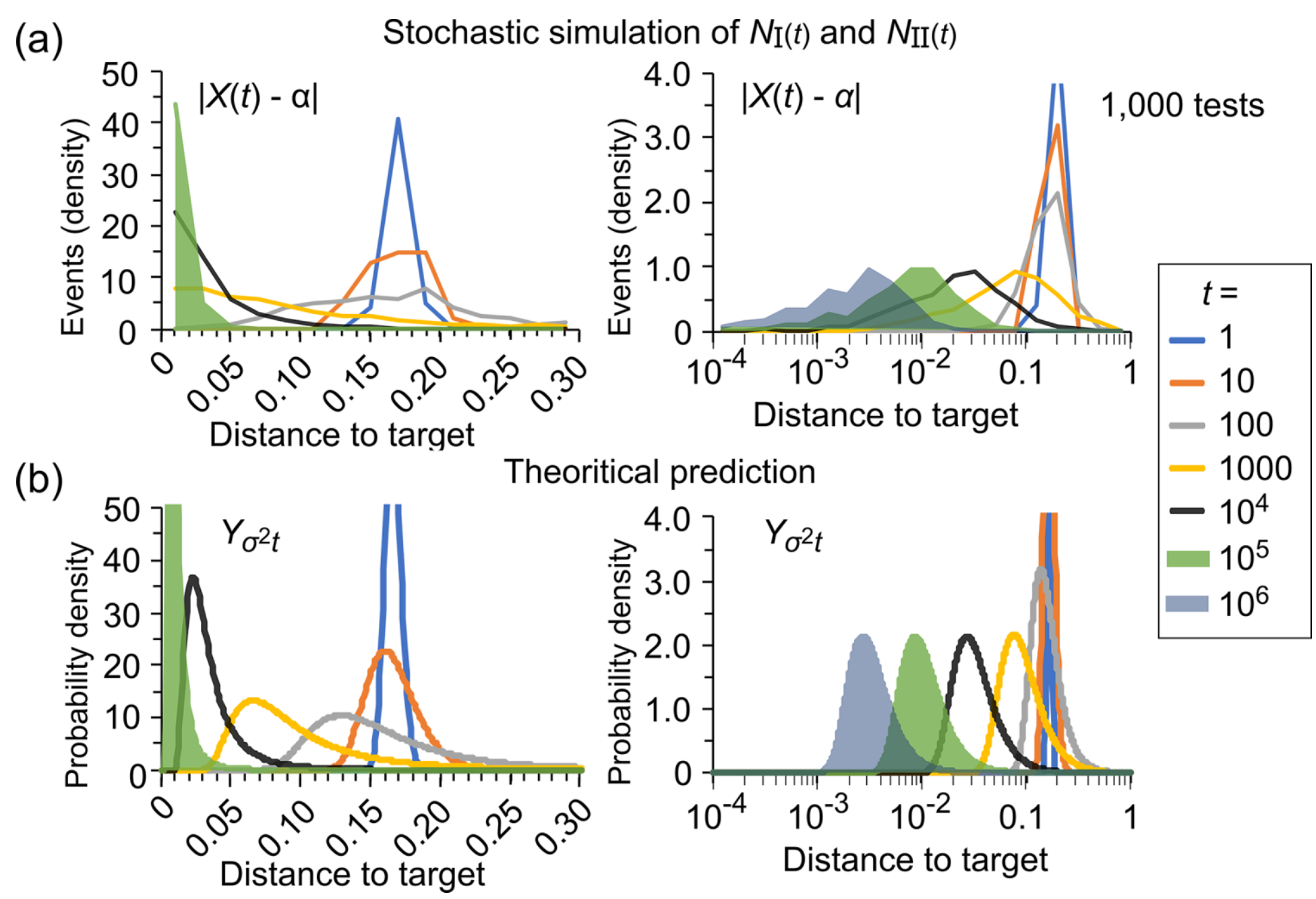

FIG. 5. Stochastic simulations and theoretical prediction in MSE regime. (a) Histograms of $|N_{\rm I}/\Sigma N_j(t) - 1/3|$ is shown, where cell numbers $N_{\rm I}(t)$ and $N_{\rm II}(t)$ change via competitive amplification with $a = 0.1$ and decay rate of 0.1 MSE. The linear-scale histogram (bin width = 0.02) shows normalized densities as event counts/1000/0.02. On a log scale, bin width = 0.2. (b) Probability densities $Y_{\sigma^2 t}$ are calculated from Eq. (21) with $a = 0.1$, $b = 0$, $\alpha = 1/3$, $k = 1$, $\sigma^2 = 4/90$, and $Y_0 = 1/6$. On a log scale, $f(x) = (\ln 10)10^x\, p(10^x, \sigma^2 t)$ is plotted.

Further, we solve the SED [Eq. (16)] under the simplified drift-free condition $b = 0$. Here, we rewrite Eq. (16) as

$$dY_t = \frac{1}{M(Y_t)} dW_t \quad \text{or} \quad dY_t = \frac{\lambda(Y_t)}{a} dW_t, \tag{17}$$

where $Y_t = |X(t/\sigma^2) - \alpha|$ denotes the distance to the target value $\alpha$. Note that $W_t$ is a Wiener process, satisfying $dt = (dW_t)^2$.

In the random walk corresponding to Fig. 2(c), $\lambda(Y_t) = akY_t$ with $0 < k \ll 1$. Substituting into Eq. (17) yields

$$dY_t = kY_t\, dW_t,$$

with the solution

$$Y_t = Y_0 \exp\left(-\frac{k^2}{2}t + kW_t\right), \tag{18}$$

where $Y_0$ is the initial distance to the target. This describes geometric Brownian motion without a drift term, a widely studied model in mathematical finance [22]. In this type of random walk, even with only a 50% chance of moving in the correct direction at each step, the distance $Y_t$ follows log-normal distributions with a constant mean and a progressively decreasing median. Notably, in Figs. 2(c) and 2(g), a small random influx $b$ introduced to prevent extinction interrupts the martingale and allows the mean to drift toward the target ratio. Thus, even without directional control, agents autonomously approach the target through repeated steps in most cases.

In the model corresponding to Fig. 2(b), $\lambda(Y_t) = akY_t^2$. Substituting into Eq. (17) gives

$$dY_t = kY_t^2\, dW_t. \tag{19}$$

Applying Itô's formula with the transformation $Z_t = 1/(kY_t)$ [see Appendix D 5], we obtain

$$dZ_t = \frac{1}{Z_t} dt - dW_t. \tag{20}$$

This is the SDE of a three-dimensional Bessel process, where $Z_t$ represents the radial component of three-dimensional Brownian motion [23]. As $t \to \infty$, it is known that particles diffuse away with $Z_t \to \infty$ almost surely, implying $Y_t \to 0$. The corresponding probability density function is

$$p(y, t|Y_0) = \frac{Y_0}{\sigma k\sqrt{2\pi t}\; y^3}\left[\exp\left(-\frac{1}{2\sigma^2 k^2 t}\left(\frac{1}{y} - \frac{1}{Y_0}\right)^2\right) - \exp\left(-\frac{1}{2\sigma^2 k^2 t}\left(\frac{1}{y} + \frac{1}{Y_0}\right)^2\right)\right]. \tag{21}$$

To verify this mathematical analysis, we compared the theoretical prediction with stochastic simulations of two-population dynamics governed by Eqs. (1) and (3). In this simulation, the cell numbers $N_{\rm I}(t)$ and $N_{\rm II}(t)$ undergo competitive amplification (with $a = 0.1$) and decay based on MSE, with $\lambda(X(t)) = 0.1[X(t) - \alpha]^2$, where the target $\alpha = 1/3$. The initial condition was set to $N_{\rm I}(0) = N_{\rm II}(0) = 18$. To avoid extinction, the cell number $N_i(t)$ was reset to one if it drew zero during the stochastic decay process. Using Eq. (12), we estimate the noise intensity $\sigma^2 = 2a\alpha(1 - \alpha) \cong 0.044$. Figure 5(a) shows the evolution of the distance $|X(t) - \alpha|$ across 1000 stochastic simulations, while Fig. 5(b) displays the corresponding theoretical distribution computed from Eq. (21). Although the simulations exhibited larger variance than predicted, the medians converged toward zero over time, requiring 100 times more repetitions to reduce the medians to 1/10. Because a small fraction with exceptionally large values is hidden in Fig. 5(b), the mean within the range [0, 1] converges to zero without broadening the distribution on a logarithmic scale. Thus, Eq. (21) captures the essential population dynamics in this MTA model.

## IV. DISCUSSION

### A. Entropy-based explanation

At first glance, it may appear paradoxical that a drift-free martingale process can generate adaptive behavior. This effect does not arise from hidden directional control, but from an entropy-based statistical principle in the Boltzmann sense: A macroscopic state compatible with larger numbers of microscopic configurations statistically dominates. Here, a microscopic configuration denotes the integer population vector $\mathbb{N}(t)$ and a macroscopic state is the population ratio $X(t) \equiv N_{\mathrm{I}}(t)/M(t)$. The number of integer states of $N_{\mathrm{I}}(t)$ compatible with a composition window $|X(t) - \alpha| \leqslant \varepsilon$ increase as the total population $M(t)$ increases, so larger populations include more microscopic states in each macroscopic state. On the other hand, compositions associated with reduced decay probabilities allow $M(t)$ to grow, whereas those with higher decay probabilities rarely sustain large populations. Consequently, accessible microscopic states become concentrated in macroscopic region associated with reduced decay. Thus, random exploration in accessible microscopic state space, which proceeds as the stochastic turnover of cells [21], yields persistent macroscopic organization without prescribed drift.

### B. Low-decay states in intestinal immunity

The regulation of intestinal immunity illustrates how low-decay regions can be defined within the MTA framework. In the CSC framework, when ROR$\gamma$t$^+$APCs activate $\mathrm{T_{reg}}$ as key regulators [4–9], they are often interpreted as selectively promoting $\mathrm{T_{reg}}$ rather than $\mathrm{T_{eff}}$. In contrast, the MTA framework proposes that a key regulator reduces the decay probability of both $\mathrm{T_{reg}}$ and $\mathrm{T_{eff}}$ populations. Supporting this view, ROR$\gamma$t$^+$ cells are essential for the formation of secondary lymphoid organs [24], enhance the survival of both $\mathrm{T_{reg}}$ and $\mathrm{T_{eff}}$ cells [25], and are more abundant in healthy intestines than in inflammatory conditions [6]. Therefore, if ROR$\gamma$t$^+$APCs—reflecting intestinal health—nonspecifically increase T cell population size, the $\mathrm{T_{reg}}/\mathrm{T_{eff}}$ ratio tends toward a balance in which ROR$\gamma$t$^+$APCs are prevalent. Importantly, this balance is naturally adaptive to context: A high $\mathrm{T_{reg}}$ ratio at homeostasis or $\mathrm{T_{eff}}$ dominance during infection both protect intestinal tissue and thereby sustain ROR$\gamma$t$^+$APC numbers. Thus, within the MTA framework, a healthy intestine is both a consequence and a cause of balanced immunity, emerging from the large number of microscopic configurations included in a proper immune balance, rather than from a precisely designed regulatory network.

### C. General structural principle of MTA

The adaptive behavior emerging in MTA [Eq. (15), $b \cong 0$] relies on four structural conditions that define a broad class of stochastic processes.

First, in most events, the stochastic variable evolves as a martingale in the absence of drift. Specifically, the expected change is zero at any interior state. In this drift-free limit, no direction in state space is intrinsically preferred.

Second, the stochastic variable is bounded. For example, the composition is confined to the finite interval [0, 1].

Third, absorbing boundary states are excluded. In the population example, the boundary corresponds to extinction of a component; however, such rare events are regularized in practice by small influx or reset rules. Mathematically, this corresponds to imposing reflecting (zero-flux) boundary conditions in the associated Fokker-Planck equation. Notably, rare influx events intermittently interrupt exact martingale behavior, allowing the ensemble mean to evolve over long timescales.

Fourth, the resistance to stochastic fluctuations is dynamically coupled to the current state. Specifically, the total population size increases in states associated with reduced decay (or enhanced survival), thereby reducing the magnitude of compositional fluctuations by each stochastic event.

Under reflecting boundaries in the drift-free limit, the stationary probability density is proportional to the square of the resistance. Consequently, an appropriate state (e.g., a balanced composition) can emerge together with a large total population size, without relying on prescribed directional control. In contrast to deterministic dynamical systems, where fixed points are defined by vanishing drift, here every interior state is drift neutral in expectation, and stabilization arises from state-dependent resistance.

### D. Biological plausibility

The present modeling framework is grounded in the fundamental characteristics of living systems. Conventional drift-driven dynamical models achieve convergence through predefined regulatory interactions, whose parameters and initial conditions are typically constrained by experimental observations. In contrast, the current model is formulated as a martingale process, in which explicit directional regulation is absent and long-term distributions exhibit only weak dependence on initial conditions (Fig. 3). Rather than encoding system-specific regulatory rules, the model is constructed from minimal features broadly associated with living systems—continuous turnover, reproduction and decay, stochasticity, and semi-isolated boundary conditions. Proliferation is modeled as amplification under mutual competition for a limited amount of growth stimulus, avoiding unlimited exponential expansion. In this sense, the biological grounding of the model lies not in parameter fitting to a particular dataset, but in the abstraction of universal biological processes, thereby reducing arbitrary or system-specific assumptions compared with conventional mechanistic models.

### E. Testable predictions

Conventional dynamical models typically predict directional changes following perturbations, particularly over short timescales. By contrast, the principal prediction of the martingale framework is statistical persistence of the present state, while individual trajectories may diverge. In fact, two qualitative empirical features are widely observed across biological systems: substantial divergence between individuals and relative temporal stability within each individual. Martingale dynamics naturally reproduce this coexistence of variability and persistence, whereas conventional drift-driven models often capture one tendency at the expense of the other. Although the martingale model does not predict precise trajectories beyond the persistence of the current state, this feature may correctly reflect the inherent uncertainty and constraints that govern future states in biological systems.

At longer timescales, the model in Fig. 2(b) indicates that recovery from catastrophic deterioration may require on the order of $10^3$–$10^4$ cell divisions. For immune cell populations in a Peyer's patch, this corresponds approximately to several days, which is of the same order as the natural course of transient infectious colitis. This comparison is intended only as an illustrative consistency check rather than a quantitative validation, since population sizes and proliferation or decay rates vary substantially across individuals and conditions.

The martingale framework yields experimentally accessible statistical predictions as follows.

Weaker directional predictability at shorter timescales. Even with complete knowledge of the present microscopic configuration, future trajectories remain uncertain, whereas macroscopic states become statistically more reproducible over longer timescales.

Diversity-robustness relationship. Macroscopic states that are more robust correspond to a larger number of compatible microscopic configurations. Such relationships may become experimentally testable through modern single-cell measurements.

Conventional drift-driven systems can be compared to engineered machines composed of identical components, while accumulated noise tends to destroy the ordered state over time. A key question is whether biological systems are organized in a similar way to such engineered machines, or instead maintain macroscopic organization through intrinsic noise and microscopic diversity. Experimental confirmation of the predicted tendencies would suggest that the MTA framework contributes to organization more substantially than conventional drift-driven models.

### F. Relation to directional control

The present modeling framework does not exclude conventional drift-driven regulatory mechanisms. In Eq. (15), explicit directional control appears as a drift term, whereas the analysis focuses primarily on the near-zero-drift regime ($b \cong 0$). Simulation and mathematical analysis (Fig. 4) show that appropriate directional regulation via $\beta$ and stochastic self-optimization via $\alpha$ can work cooperatively, especially when $\beta \cong \alpha$. For example, artificial learning algorithms control both the direction and the extent of parameter updates [11,12]. The framework is therefore intended to complement, rather than replace, mechanistic dynamical models.

Martingale-based self-optimizing processes may be underappreciated in biological modeling, partly due to their inherent stochasticity and causal ambiguity [13,14]. Directional regulatory mechanisms are more amenable to mechanistic identification and to experimental or therapeutic intervention. Consequently, biological studies often emphasize specific, reproducible regulation of particular cell types. In this view, high specificity tends to be equated with biological importance, whereas nonspecific, martingale-type effects may remain overlooked or unpublished. However, such a bias may partly contribute to the reproducibility crisis in biological research. The present framework [Eq. (15)] mathematically describes how directional control and resistance to change can operate cooperatively.

### G. Interdisciplinary relevance

The present MTA framework is conceptually related to generative thermodynamic computing. Recent studies have shown that structure can arise through physical dynamics from noise encoded in a system's energy landscape [26]. Our model, grounded in biological systems, generates appropriate population balance through random walk with state-dependent resistance to change, reflecting a macroscopic fitness landscape. Both systems operate without active denoising, which is energetically costly in digital computers. Thus, martingale-based biological regulation would be of interest not only in theoretical biology but also in broader studies of stochastic computation and nonequilibrium statistical physics.

## V. CONCLUSIONS

We propose a theory of biological regulation based on the convergence of stochastic processes, rather than the traditional notion of fixed-point stability in drift-driven dynamical systems. Specifically, we apply Doob's martingale convergence theorem [18] to population balance, assuming that proliferation and death occur stochastically, without component-specific regulation. When cell decay rates are low, the total population size increases, and the balance becomes resistant to perturbation. Consequently, appropriate population compositions—in practice synonymous with low-decay states—emerge naturally through stochastic turnover. We refer to this principle as resistance-mediated adaptation, in which adaptive organization arises through regulation of state-dependent resistance to stochastic change rather than through explicit directional control.

## ACKNOWLEDGMENTS

T.Y. acknowledges the Tokushukai Medical Group for financial support, Tetsuya J. Kobayashi for helpful discussions, and the use of DeepSeek-R1 for mathematical analyses and ChatGPT-5 for English editing.

## DATA AVAILABILITY

The data that support the findings of this article are openly available [27].

## APPENDIX A: COMPETITION IN PROLIFERATION

In the proliferation process in the main model, we consider a competitive situation in which the total population grows with a constant probability. Although this assumption may appear unconventional when interpreted as a direct proliferation model, it is natural when understood as a resource-limited growth process. Many birth-death models of population dynamics assume that the growth rate increases with population size, presuming that triggers and nutrients for growth are abundant. We instead assume that the triggering signal for cell division is scarce relative to the number of cells and is immediately consumed upon use. In our previous study on epigenetic gene regulation [16], we proposed a model in which a histone acetylase attaches to and acetylates a gene locus depending on the acetylation ratio among candidate genes. The enzyme loses its activity once it acetylates a site. Similarly, we consider a biological situation in which a growth signal stimulates a cell in the population and becomes inactivated immediately after use. This assumption prevents an unrealistic exponential increase in population size. Therefore, assuming a constant total population growth rate is more plausible than models that postulate a linear increase in growth rate with population size, at least in certain biological contexts or in a short period.

In a generalized form of our model, we assess the effect of growth rate. Instead of Eq. (1), we here assume that each cell divides with probability $g(M(t))\tau$ during the interval $[t, t+\tau)$, where the total cell number is $M(t) \equiv \sum N_j(t)$. The number of type-$i$ cells after division follows a binomial distribution:

$$P(N_i(t+\tau) = n + N_i(t)|\mathbb{N}(t)) = \binom{N_i(t)}{n}[g(M(t))\tau]^n\,[1-g(M(t))\tau]^{N_i(t)-n}, \quad \text{(A1)}$$

where $n \in \{0, 1, \ldots, N_i(t)\}$. The expected growth in this stochastic process is $\langle N_i(t+\tau) - N_i(t)\rangle = g(M(t))N_i(t)\tau$. Therefore, $g(M(t)) = a/M(t)$ is almost equivalent to Eq. (1). We performed stochastic simulations using various forms of $g(M(t))$ in Fig. 6. As in the MSE model [Fig. 2(b)], initial values $N_{\rm I}(0) = N_{\rm II}(0) = 1$, $b_i = 0.0005$, and the decay probability is the larger of $10^{-4}$ and $0.1[X(t) - \alpha]^2$, with $\alpha = 1/3$, which is set as a biologically appropriate target composition.

When $g(M(t))$ is constant, cell number increased exponentially [Figs. 6(a) and 6(b)]. In each simulation, the population ratio $X(t) \equiv N_{\rm I}(t)/M(t)$ stabilized at a stochastically determined value. When the growth rate was set low, the population ratio $X(t)$ was slightly skewed toward $\alpha = 1/3$ [Fig. 6(b)]. In any case, such exponentially rapid increases in cell number are unrealistic in biological contexts.

To prevent unbounded proliferation, we impose a maximum on the total population growth rate, $\langle M(t+\tau) - M(t)\rangle/\tau = g(M(t))M(t)$. Following the Michaelis-Menten kinetics of biochemical reactions, we set $g(M(t)) = a/[K_{\rm m} + M(t)]$. Figures 6(c)–6(h) show simulation results obtained with $a = 0.1$ or 1 and $K_m = 100$, 10, or 1. As expected, these settings prevented excessive increases in population size under conditions where the decay probability has its minimal value $10^{-4}$. When $K_m$ was large, however, the growth rate at small population sizes became too low. For example, with $a = 0.1$ and $K_m = 100$, either $N_{\rm I}(t)$ or $N_{\rm II}(t)$ was zero at $t = 10^5$ in 938 of 1000 simulations [Fig. 6(c)]. On the other hand, at a higher maximal growth rate ($a = 1$), proliferation occurred after a long initial lag, and the evolving cell populations tended to have a composition with $X(t = 10^5)$ near $\alpha = 1/3$ [Fig. 6(f)].

Interestingly, lower $K_m$ values yielded more favorable results. When $K_m = 1$, the resulting nearly constant growth rate alleviated both issues: the extremely low growth rate at small population sizes and the excessively high growth rate at large population sizes. This resulted in an appropriate population size and composition [Figs. 6(e) and 6(h)]. When $K_m = 0$ [Figs. 6(k) and 6(n)], $g(M(t)) = a/M(t)$, which is nearly equivalent to the MTA model in the main text [Eq. (1) and Fig. 2(b)]. In this formulation, the parameter $a$ corresponds to the maximal reaction rate in Michaelis-Menten kinetics. The Michaelis constant $K_m$ generally represents the inverse of the affinity between an enzyme and its substrate. Thus, a small $K_m$ implies high affinity between cells and the growth signal. Under such conditions, each available growth signal—provided as an influx—stimulates a cell division event and is immediately inactivated after use.

Population dynamics can also be modeled as spatially constrained colonies. In a $d$-dimensional colony, the growth rate may be limited by its $(d-1)$-dimensional surface area, which governs access to growth signals or nutrients. To evaluate this effect, we consider $g(M(t)) = a/M(t)^{1/d}$ as the proliferation function in Figs. 6(i)–6(n). Simulation results show that, when $a = 0.1$ and $d = 2$, the population size and composition are regulated to some extent [Fig. 6(j)]. When $d = 1$, $g(M(t)) = a/M(t)$, which is nearly equivalent to the setting in Eq. (1). In this one-dimensional case [Figs. 6(k) and 6(n)]—corresponding to a suspended or filamentous colony where the access to growth signals by the colony is constant and all cells in the colony mutually compete for the signal—the population size and composition are autonomously regulated through the mechanism described by the MTA model in the main text.

These results collectively indicate that our formulation $g(M(t)) = a/M(t)$ represents a biologically plausible growth process in a suspended colony with enzymatic saturation. Although the proliferation probability of each cell decreases inversely with total population size, this behavior naturally arises from surface-limited growth and Michaelis-Menten-type kinetics. For example, in intestinal immunity, professional APCs constitute relatively small population compared with T lymphocytes, yet effectively control their proliferation. These findings support the biological relevance of our model, in which cells proliferate under mutual competition.

## APPENDIX B: WGM RELEVANT TO MSE AND TO BIOLOGICAL SYSTEMS

WGM stands for the weighted geometric mean of the relative appropriateness across all cell types: $\prod_j (\frac{\alpha_j}{X_j})^{X_j}$, where $X_j$ is the composition of type-$j$ cells and $\alpha_j$ is its target proportion. Since $\Sigma X_j = 1$ and $\Sigma\alpha_j = 1$, WGM ranges from 0 to 1, reaching 1 when $X_j = \alpha_j$ for all $j$.

In the MTA framework, the compositions of two cell populations showed similar distributions whether the decay function $\lambda(X(t))$ was defined by MSE or by WGM [Figs. 2(b), 2(d), and 2(g)]. The following approximation explains this similarity. When $X(t)$ is transformed to the deviation from the

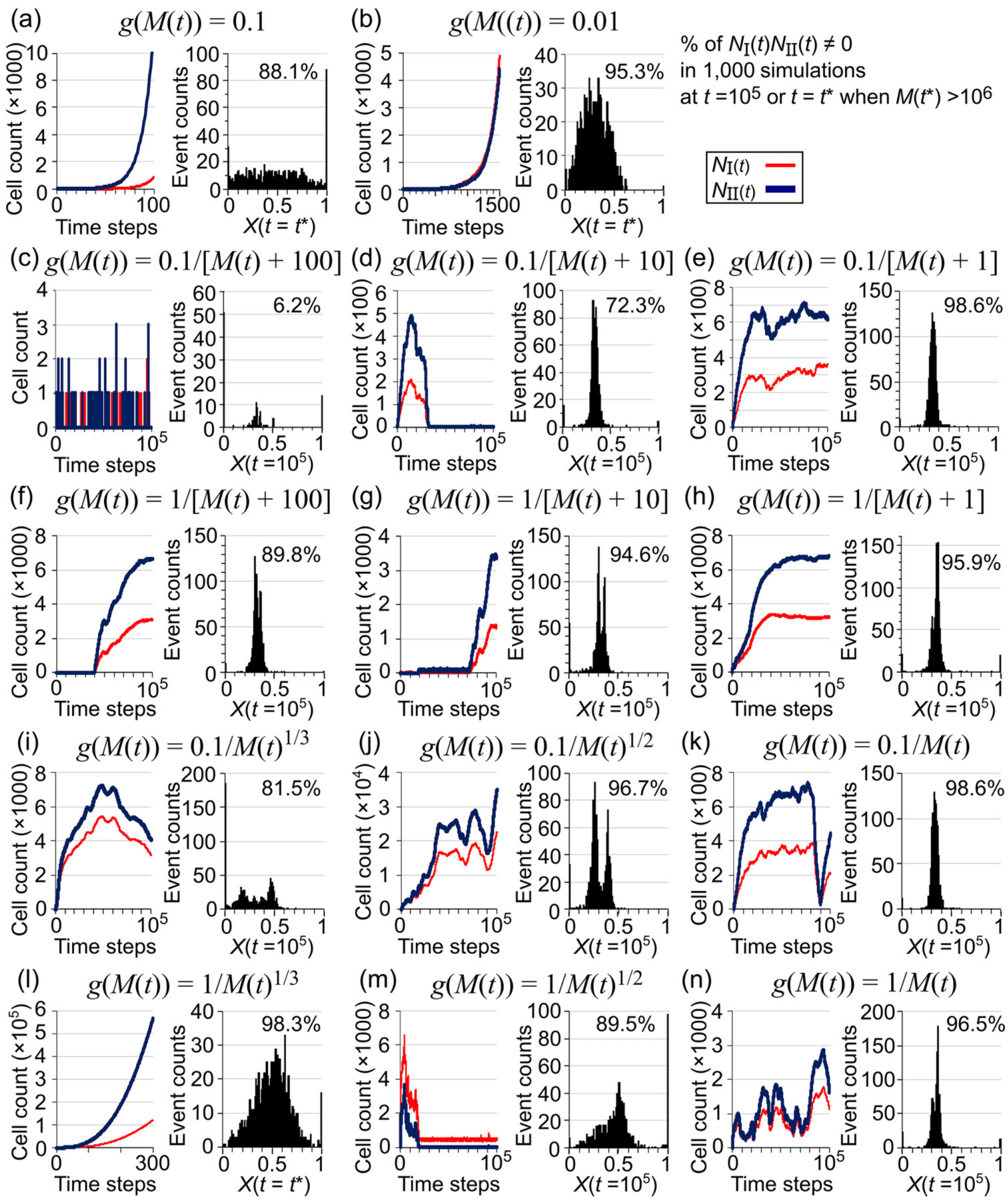

FIG. 6. Simulations under different growth conditions. The cell numbers $N_{\rm I}(t)$ and $N_{\rm II}(t)$ increase via proliferation with probability $g(M(t))$ and influx $b_i = 0.0005$, and decay with an MSE-based probability given by the larger of $0.1[X(t) - 1/3]^2$ and $10^{-4}$, at each time step. In contrast to the main model where only one cell proliferates per step, here each cell divides independently with probability $g(M(t))$, following a binomial process similar to that used for decay. Population dynamics over time in a simulation and histograms of $X(t)$ at the end of the simulation [$t = 10^5$ or $t^*$ when $M(t^*) > 10^6$] are shown under each growth condition. Histograms (bin width = 0.01) show event counts from 1000 simulations, excluding cases with $M(t = 10^5) = 0$. The percentage of cases where both $N_{\rm I}(t)$ and $N_{\rm II}(t)$ are nonzero is inserted in each histogram. The setting in (k) is nearly equivalent to that in Fig. 2(b). $M(t) \equiv \sum N_j(t)$. $X(t) \equiv N_{\rm I}(t)/M(t)$. (a) $g(M(t)) = 0.1$, (b) $g(M(t)) = 0.01$, (c) $g(M(t)) = 0.1/[M(t) + 100]$, (d) $g(M(t)) = 0.1/[M(t) + 10]$, (e) $g(M(t)) = 0.1/[M(t) + 1]$, (f) $g(M(t)) = 1/[M(t) + 100]$, (g) $g(M(t)) = 1/[M(t) + 10]$, (h) $g(M(t)) = 1/[M(t) + 1]$, (i) $g(M(t)) = 0.1/M(t)^{1/3}$, (j) $g(M(t)) = 0.1/M(t)^{1/2}$, (k) $g(M(t)) = 0.1/M(t)$, (l) $g(M(t)) = 1/M(t)^{1/3}$, (m) $g(M(t)) = 1/M(t)^{1/2}$, (n) $g(M(t)) = 1/M(t)$.

target, $Y(t) \equiv \alpha - X(t)$, and $Y(t) \cong 0$, the following holds:

$$\begin{aligned}
\lambda(X(t)) &= 1 - \mathrm{WGM}^k = 1 - \left[\frac{\alpha}{X(t)}\right]^{kX(t)}\left[\frac{1-\alpha}{1-X(t)}\right]^{k[1-X(t)]} = 1 - \left[\frac{\alpha}{\alpha - Y(t)}\right]^{k[\alpha - Y(t)]}\left[\frac{1-\alpha}{1-\alpha+Y(t)}\right]^{k[1-\alpha+Y(t)]} \\
&= 1 - \left[1 + \frac{Y(t)}{\alpha - Y(t)}\right]^{k[\alpha - Y(t)]}\left[1 - \frac{Y(t)}{1-\alpha+Y(t)}\right]^{k[1-\alpha+Y(t)]} \\
&\cong 1 - [1 + kY(t)][1 - kY(t)] = [kY(t)]^2 = k^2[\alpha - X(t)]^2 \therefore 1 - \mathrm{WGM}^k \cong k^2\mathrm{MSE},\ if\ X(t) \cong \alpha.
\end{aligned} \tag{B1}$$

As shown in Eq. (4), WGM is mathematically equivalent to $e^{-D_{\mathrm{KL}}}$, where $D_{\mathrm{KL}}$ denotes the Kullback-Leibler divergence, an information-theoretic measure of the deviation of the current population composition from its target. Accordingly, WGM is related to both MSE and $D_{\mathrm{KL}}$, which are widely used in artificial systems for evaluating and correcting errors.

The WGM′ model also provides a biologically plausible estimate of fitness. Each cell may estimate the ratio of its target proportion to its current fraction, $\alpha_i/X_i(t)$, and the WGM of these ratios across all types determines the survival probability of cells. Biologically, $\alpha_i/X_i(t)$ can be interpreted as a measure of demand per cell. When these demands are pooled across cell types, the WGM yields a global estimate of survival probability. Therefore, decay probability functions based on WGM, MSE, or $D_{\mathrm{KL}}$ may have biological relevance in the MTA framework, effectively favoring compositions that minimize the information distance between the current state and functional demand.

## APPENDIX C: NOISE EFFECTS IN MTA AND CSC

We describe the asymptotic behavior of $X(t)$ based on Eq. (14). When the decay function $\lambda(X(t))$ is modeled as MSE, i.e., $\lambda(X(t)) = [X(t)-\alpha]^2$, Eq. (14) becomes

$$\frac{dX(t)}{dt} = -\frac{b}{a+b}\left[X(t)-\beta-\frac{\sigma}{b}\eta(t)\right][X(t)-\alpha]^2. \quad \text{(C1)}$$

Temporarily fixing the noise value, we can plot $dX(t)/dt$ as a function of $X(t)$ (Fig. 4).

In the CSC regime, where the influx parameter $b \gg a$, the noise term $\sigma\eta(t)/b$ becomes negligible. Under this condition, $\beta + \sigma\eta(t)/b$ acts as a stable fixed point toward which $X(t)$ converges. The target value $\alpha$ instead acts as a saddlelike point. Specifically, if $\alpha < \beta$, then $X(t) < \alpha$ increases toward $\alpha$, with the rate of change slowing down as it approaches $\alpha$ (solid curve in Fig. 4). However, once $X(t) > \alpha$, $X(t)$ accelerates away from $\alpha$ toward $\beta + \sigma\eta(t)/b$. Conversely, if $\alpha > \beta$, then $X(t) > \alpha$ decreases toward $\beta + \sigma\eta(t)/b$, again with transient slowing down near $\alpha$ (dotted curve in Fig. 4). Thus, in CSC, $X(t)$ tends to stabilize around $\beta$, with only limited fluctuations.

In contrast, in the MTA regime where $b \cong 0$, the noise term $\sigma\eta(t)/b$ becomes large in magnitude relative to both $\beta$ and $X(t)$. As a result, the putative fixed point $\beta + \sigma\eta(t)/b$ becomes unstable, as illustrated by two curves in Fig. 4. The sign of $dX(t)/dt$ fluctuates depending on the instantaneous sign of $\eta(t)$, resulting in frequent reversals in the direction of change of $X(t)$. Consequently, $X(t)$ behaves as a type of random walk with step lengths that decrease near $\alpha$. Due to critical slowing down around the saddle-like point at $\alpha$, along with frequent stochastic direction reversals, $X(t)$ tends to stay near $\alpha$ for prolonged periods—consistent with both simulation results and mathematical analysis.

Notably, whereas CSC stabilizes $X(t)$ near $\beta$ by suppressing noise, MTA achieves regulation near $\alpha$ through noise-driven fluctuations. This noise-utilizing strategy resembles mechanisms proposed in thermodynamic computing [26] and may help explain the remarkable efficiency of biological brains, compared with artificial systems that require external energy to suppress thermal noise.

## APPENDIX D: DERIVATION OF SDE

We provide details of Eqs. (5), (6), (8), (10), (11), (20), and (21) presented in the main text.

### 1. Equation (5): $\frac{\langle N_i(t+\tau)-N_i(t)\rangle}{\tau} = a\frac{N_i(t)}{\sum_j N_j(t)} - \lambda(\mathbb{N}(t))\,N_i(t) + b_i$

When the time interval $\tau$ is sufficiently small, the three stochastic reactions described by Eqs. (1)–(3) occur mutually exclusively. The expected change in $N_i(t)$ over the interval $\tau$ is given by Eq. (5). We also obtain the variance [denoted by $V(\cdot)$] of the change in cell number $N_i(t)$:

$$\begin{aligned} V(N_i(t+\tau)-N_i(t)) &= \frac{aN_i(t)\tau}{\sum_j N_j(t)}\left[1-\frac{aN_i(t)\tau}{\sum_j N_j(t)}\right] + \lambda(\mathbb{N}(t))\tau[1-\lambda(\mathbb{N}(t))\tau]\,N_i(t) + b_i\tau(1-b_i\tau) \\ &\cong \left[a\frac{N_i(t)}{\sum_j N_j(t)} + \lambda(\mathbb{N}(t))N_i(t) + b_i\right]\tau. \end{aligned} \quad \text{(D1)}$$

Thus, the SDE for the cell number becomes

$$\frac{dN_i(t)}{dt} = a\frac{N_i(t)}{\sum_j N_j(t)} - \lambda(\mathbb{N}(t))\,N_i(t) + b_i + \eta(t)\sqrt{a\frac{N_i(t)}{\sum_j N_j(t)} + \lambda(\mathbb{N}(t))N_i(t) + b_i}, \quad \text{(D2)}$$

where $\eta(t)$ is temporally uncorrelated Gaussian white noise.

### 2. Equation (6): $\frac{\langle M(t+\tau)-M(t)\rangle}{\tau} = a - \lambda(X(t))M(t) + b$

For small $\tau$, at most one cell increases or decreases during the infinitesimal interval $[t, t+\tau)$. Summing Eq. (5) over all cell types yields Eq. (6), which describes the expected change in the total cell number $M(t) \equiv \sum N_j(t)$. While the decay rate can generally be written as $\lambda(\mathbb{N}(t)) = \lambda(X(t),\, M(t))$, we assume throughout our models that the decay probability per cell depends only on the composition $X(t)$ and is independent of the total population size $M(t)$. Under this assumption, $\lambda(\mathbb{N}(t))$ is rewritten as $\lambda(X(t))$. The variance of the change in the total cell number is

$$V(M(t+\tau)-M(t)) = a\tau(1-a\tau)+\lambda(X(t))\tau[1-\lambda(X(t))\tau]M(t)+b\tau(1-b\tau) \cong [a+\lambda(X(t))M(t)+b]\tau. \tag{D3}$$

Thus, the SDE for the total cell number becomes

$$\frac{dM(t)}{dt} = a - \lambda(X(t))\,M(t)+b+\eta(t)\sqrt{a+\lambda(X(t))M(t)+b}. \tag{D4}$$

### 3. Equation (8): $\frac{\langle dX(t)\rangle}{dt} = \frac{b}{M(t)+1}[\beta - X(t)]$

We derive the change in composition $X(t)$, from the change in cell number $\mathbb{N}(t)$. We assume that only one cell increases or decreases during a small interval $\tau$, which holds when $\lambda(X(t))M(t)\tau \ll 1$ and $(a+b)\tau \ll 1$, excluding the dynamics during a catastrophic phase.

In the cell increase processes given by Eqs. (1) and (2), a type-I cell is added with probability $[aX(t)+b_{\mathrm{I}}]\tau$. The change in $X(t)$ is

$$\begin{aligned} X(t+\tau)-X(t) &= \frac{N_{\mathrm{I}}(t)+1}{N_{\mathrm{I}}(t)+N_{\mathrm{II}}(t)+1} - \frac{N_{\mathrm{I}}(t)}{N_{\mathrm{I}}(t)+N_{\mathrm{II}}(t)} \\ &= \frac{N_{\mathrm{I}}(t)^2+N_{\mathrm{I}}(t)N_{\mathrm{II}}(t)+N_{\mathrm{I}}(t)+N_{\mathrm{II}}(t)-[N_{\mathrm{I}}(t)^2+N_{\mathrm{I}}(t)N_{\mathrm{II}}(t)+N_{\mathrm{I}}(t)]}{[N_{\mathrm{I}}(t)+N_{\mathrm{II}}(t)+1][N_{\mathrm{I}}(t)+N_{\mathrm{II}}(t)]} \\ &= \frac{N_{\mathrm{II}}(t)}{[N_{\mathrm{I}}(t)+N_{\mathrm{II}}(t)+1][N_{\mathrm{I}}(t)+N_{\mathrm{II}}(t)]}\ \therefore X(t+\tau)-X(t) = \frac{1-X(t)}{M(t)+1}. \end{aligned} \tag{D5}$$

A type-II cell is added with probability $\{a[1-X(t)]+b_{\mathrm{II}}\}\tau$. The change in $X(t)$ is

$$\begin{aligned} X(t+\tau)-X(t) &= \frac{N_{\mathrm{I}}(t)}{N_{\mathrm{I}}(t)+N_{\mathrm{II}}(t)+1} - \frac{N_{\mathrm{I}}(t)}{N_{\mathrm{I}}(t)+N_{\mathrm{II}}(t)} \\ &= \frac{N_{\mathrm{I}}(t)^2+N_{\mathrm{I}}(t)N_{\mathrm{II}}(t)-[N_{\mathrm{I}}(t)^2+N_{\mathrm{I}}(t)N_{\mathrm{II}}(t)+N_{\mathrm{I}}(t)]}{[N_{\mathrm{I}}(t)+N_{\mathrm{II}}(t)+1][N_{\mathrm{I}}(t)+N_{\mathrm{II}}(t)]} \\ &= \frac{-N_{\mathrm{I}}(t)}{[N_{\mathrm{I}}(t)+N_{\mathrm{II}}(t)+1][N_{\mathrm{I}}(t)+N_{\mathrm{II}}(t)]}\ \therefore X(t+\tau)-X(t) = -\frac{X(t)}{M(t)+1}. \end{aligned} \tag{D6}$$

In the cell decay process given by Eq. (3), one type-II cell is removed with probability $\lambda(X(t))N_{\mathrm{II}}(t)\tau = [1-X(t)]\,\lambda(X(t))M(t)\tau$ during the infinitesimal interval $[t, t+\tau)$. The change in $X(t)$ is

$$\begin{aligned} X(t+\tau)-X(t) &= \frac{N_{\mathrm{I}}(t)}{N_{\mathrm{I}}(t)+N_{\mathrm{II}}(t)-1} - \frac{N_{\mathrm{I}}(t)}{N_{\mathrm{I}}(t)+N_{\mathrm{II}}(t)} \\ &= \frac{N_{\mathrm{I}}(t)^2+N_{\mathrm{I}}(t)N_{\mathrm{II}}(t)-[N_{\mathrm{I}}(t)^2+N_{\mathrm{I}}(t)N_{\mathrm{II}}(t)-N_{\mathrm{I}}(t)]}{[N_{\mathrm{I}}(t)+N_{\mathrm{II}}(t)-1][N_{\mathrm{I}}(t)+N_{\mathrm{II}}(t)]} \\ &= \frac{N_{\mathrm{I}}(t)}{[N_{\mathrm{I}}(t)+N_{\mathrm{II}}(t)-1][N_{\mathrm{I}}(t)+N_{\mathrm{II}}(t)]}\ \therefore X(t+\tau)-X(t) = \frac{X(t)}{M(t)-1}. \end{aligned} \tag{D7}$$

A type-I cell is removed with probability $\lambda(X(t))N_{\mathrm{I}}(t)\tau = X(t)\lambda(X(t))M(t)\tau$. The change in $X(t)$ is

$$\begin{aligned} X(t+\tau)-X(t) &= \frac{N_{\mathrm{I}}(t)-1}{N_{\mathrm{I}}(t)+N_{\mathrm{II}}(t)-1} - \frac{N_{\mathrm{I}}(t)}{N_{\mathrm{I}}(t)+N_{\mathrm{II}}(t)} \\ &= \frac{N_{\mathrm{I}}(t)^2+N_{\mathrm{I}}(t)N_{\mathrm{II}}(t)-N_{\mathrm{I}}(t)-N_{\mathrm{II}}(t)-[N_{\mathrm{I}}(t)^2+N_{\mathrm{I}}(t)N_{\mathrm{II}}(t)-N_{\mathrm{I}}(t)]}{[N_{\mathrm{I}}(t)+N_{\mathrm{II}}(t)-1][N_{\mathrm{I}}(t)+N_{\mathrm{II}}(t)]} \\ &= \frac{-N_{\mathrm{II}}(t)}{[N_{\mathrm{I}}(t)+N_{\mathrm{II}}(t)-1][N_{\mathrm{I}}(t)+N_{\mathrm{II}}(t)]}\ \therefore X(t+\tau)-X(t) = -\frac{1-X(t)}{M(t)-1}. \end{aligned} \tag{D8}$$

By combining Eqs. (D5)–(D8) under the assumption that each event occurs mutually exclusively within the infinitesimal interval $[t, t+\tau)$, the expected change in $X(t)$ is given by

$$\begin{aligned} \langle X(t+\tau)-X(t)\rangle &= [aX(t)+b_{\mathrm{I}}]\tau\frac{1-X(t)}{M(t)+1} - \{a[1-X(t)]+b_{\mathrm{II}}\}\tau\frac{X(t)}{M(t)+1} \\ &\quad + \lambda(X(t))M(t)[1-X(t)]\tau\frac{X(t)}{M(t)-1} - \lambda(X(t))M(t)X(t)\tau\frac{1-X(t)}{M(t)-1} \\ &= \frac{b_{\mathrm{I}}[1-X(t)]-b_{\mathrm{II}}X(t)}{M(t)+1}\tau = \frac{b\beta[1-X(t)]-b(1-\beta)X(t)}{M(t)+1}\tau = \frac{b[\beta-X(t)]}{M(t)+1}\tau, \end{aligned} \tag{D9}$$

where $b \equiv b_{\rm I} + b_{\rm II}$ and $\beta \equiv b_{\rm I}/b$.

#### 4. Equation (10): $\frac{V(dX(t))}{dt} \cong \frac{2aX(t)[1-X(t)]+b[X(t)-2\beta X(t)+\beta]}{M(t)^2}$

Using Eqs. (D5)–(D8), the variance of $X(t)$ change over $\tau$ is

$$\begin{aligned}
&V(X(t+\tau)-X(t)) + \langle X(t+\tau)-X(t)\rangle^2\\
&\quad= [aX(t)+b\beta]\tau\frac{[1-X(t)]^2}{[M(t)+1]^2} + \{a[1-X(t)]+b(1-\beta)\}\tau\frac{X(t)^2}{[M(t)+1]^2}\\
&\qquad+\lambda(X(t))M(t)[1-X(t)]\tau\frac{X(t)^2}{[M(t)-1]^2} + \lambda(X(t))M(t)X(t)\tau\frac{[1-X(t)]^2}{[M(t)-1]^2}\\
&\quad= X(t)[1-X(t)]\left\{\frac{a}{[M(t)+1]^2}+\frac{\lambda(X(t))M(t)}{[M(t)-1]^2}\right\}\tau + \frac{b\beta[1-X(t)]^2+b(1-\beta)X(t)^2}{[M(t)+1]^2}\tau\\
&\quad= X(t)[1-X(t)]\left\{\frac{a}{[M(t)+1]^2}+\frac{\lambda(X(t))M(t)}{[M(t)-1]^2}\right\}\tau + \frac{b\big[X(t)^2-2\beta X(t)+\beta\big]}{[M(t)+1]^2}\tau.
\end{aligned}\tag{D10}$$

Since the squared expectation is negligible ($\sim\tau^2$), this gives Eq. (9).
Assuming $M(t) \gg 1$ and $\lambda(X(t))M(t) \cong a+b$, approximation gives Eq. (10):

$$\begin{aligned}
\frac{V(X(t+\tau)-X(t))}{\tau} &\cong \frac{X(t)[1-X(t)][a+\lambda(X(t))M(t)]+b\big[X(t)^2-2\beta X(t)+\beta\big]}{M(t)^2}\\
&\cong \frac{X(t)[1-X(t)](a+a+b)+b\big[X(t)^2-2\beta X(t)+\beta\big]}{M(t)^2}\\
&= \frac{2aX(t)[1-X(t)]+b[X(t)-2\beta X(t)+\beta]}{M(t)^2}.
\end{aligned}\tag{D11}$$

Here, we define a noise intensity function as

$$\sigma^2(X(t)) \equiv V(X(t+\tau)-X(t))\frac{M(t)^2}{\tau} = 2aX(t)[1-X(t)] + b\{(1-\beta)X(t)+\beta[1-X(t)]\}.\tag{D12}$$

Since $X(t)$ is within the range [0, 1], the first term ranges in $[0, a/2]$ and the second term in $[b\beta,\ b(1-\beta)]$. Therefore, the minimum of $\sigma^2(X(t))$ is greater than $\min\{b\beta, b(1-\beta)\}$, and the maximum is less than $a/2+\max\{b\beta, b(1-\beta)\}$. For example, in the MTA regime, where $a=0.1$, $b=0.001$, $\beta=0.5$, $\min[\sigma^2(X(t))]=0.0005$ [when $X(t)=0$ or 1], and $\max[\sigma^2(X(t))]=0.0505$ [when $X(t)=0.5$]. In the CSC regime, where $a=0$, $b=0.1$, $\beta=0.33$, $\min[\sigma^2(X(t))]=0.033$ [when $X(t)=0$], and $\max[\sigma^2(X(t))]=0.066$ [when $X(t)=1$]. If $b=0$, $\sigma^2(X(t))=0$ when $X(t)=0$ or 1, implying extinction of a cell type or absorbing boundary. Therefore, when $X(t)\cong 1/2$, $\sigma^2(X(t))$ is relatively stable. In contrast, the total population size $M(t)$ can vary widely when $X(t)\cong\alpha$ and $\lambda(X(t)\cong\alpha)\cong 0$. Specifically, in the MSE model, where $\lambda(X(t))=k[X(t)-\alpha]^2$, $M(t)=[a+b]/\lambda(X(t))$ varies in a wide range while $X(t)\cong\alpha$, because $\lambda(X(t)\cong\alpha)$ yields nearly zero. Taken together, under the conditions setting of $0<a+b\ll 1$, $M(t)X(t)\gg 1$, and $M(t)(1-X(t))\gg 1$, the variance of $X(t)$ change is predominantly determined by $M(t)$. Thus, we simplify $V(dX(t))/dt \cong \sigma^2/M(t)^2$ using a constant parameter $\sigma^2$. When $X(t)\cong\alpha$, in either MTA ($a\gg b$) or CSC ($\beta\cong\alpha$) regimes, substituting to Eq. (D12) gives $\sigma^2(\alpha)=2a\alpha(1-\alpha)+b(\alpha+\beta-2\alpha\beta)$. This corresponds to Eq. (12). When $X(t)\cong\beta$ under CSC, substituting to Eq. (D12) gives $\sigma^2(\beta)=2a\beta(1-\beta)+b((1-\beta)\beta+\beta(1-\beta))=2(a+b)\beta(1-\beta)$.

#### 5. Equation (21): $p(y,t|Y_0) = \frac{Y_0}{\sigma k\sqrt{2\pi t}\ y^3}[\exp(-\frac{1}{2\sigma^2k^2t}(\frac{1}{y}-\frac{1}{Y_0})^2) - \exp(-\frac{1}{2\sigma^2k^2t}(\frac{1}{y}+\frac{1}{Y_0})^2)]$

To solve $dY_t = kY_t{}^2\ dW_t$ [Eq. (19)], we apply Itô's formula. For a general SDE of the form $dX_t=\mu_t\ dt+\sigma_t\ dW_t$, Itô's formula states

$$df(t,x) = \left(\frac{\partial f}{\partial t}+\mu_t\frac{\partial f}{\partial x}+\frac{\sigma_t{}^2}{2}\frac{\partial^2 f}{\partial x^2}\right)dt + \sigma_t\frac{\partial f}{\partial x}dW_t.\tag{D13}$$

In the case of Eq. (19), we set $\mu_t=0$ and $\sigma_t=kY_t^2$. Let us define the transformation: $Z_t=\frac{1}{k\ Y_t}$.
Then, the derivatives are $\frac{\partial Z_t}{\partial y}=-\frac{1}{k\ Y_t{}^2}$, $\frac{\partial^2 Z_t}{\partial y^2}=\frac{2}{k\ Y_t{}^3}$, $\frac{\partial Z_t}{\partial t}=0$.
Substituting these into Eq. (D13) gives Eq. (20):

$$dZ_t = \left(\frac{\partial Z_t}{\partial t}+0\ \frac{\partial Z_t}{\partial y}+\frac{k^2\ Y_t{}^4}{2}\frac{\partial^2 Z_t}{\partial y^2}\right)dt + kY_t{}^2\frac{\partial Z_t}{\partial y}dW_t = \frac{k^2Y_t{}^4}{2}\frac{2}{k\ Y_t{}^3}dt - kY_t{}^2\frac{1}{k\ Y_t{}^2}dW_t = kY_t dt - dW_t = \frac{1}{Z_t}dt - dW_t.$$

This Eq. (20) is known as the three-dimensional Bessel process [23]. Its probability density function is given by

$$p_Z(z,t|Z_0) = \frac{1}{t}\frac{z\sqrt{z}}{\sqrt{Z_0}}\exp\left(-\frac{z^2+{Z_0}^2}{2t}\right)\left(\frac{\exp\left(\frac{Z_0 z}{t}\right)-\exp\left(-\frac{Z_0 z}{t}\right)}{\sqrt{2\pi\frac{Z_0 z}{t}}}\right) = \frac{z}{Z_0\sqrt{2\pi t}}\left(\exp\left(-\frac{(z-Z_0)^2}{2t}\right)-\exp\left(-\frac{(z+Z_0)^2}{2t}\right)\right), \tag{D14}$$

where $Z_0$ is the initial value of $Z_t$. To match the discrete time steps in the simulation, we replace $t$ with $\sigma^2 t$, and change variables substituting $z = 1/(ky)$. Then, we obtain the probability density function for $Y(t) = |X(t)-\alpha|$:

$$\begin{aligned}
p_Y(y,t|Y_0) &= p_z\left(\frac{1}{ky},\sigma^2 t|\frac{1}{kY_0}\right)\cdot\left|\frac{dz}{dy}\right| = p_z\left(\frac{1}{ky},\sigma^2 t|\frac{1}{kY_0}\right)\cdot\frac{1}{ky^2}\\
&= \frac{\frac{1}{ky}}{\frac{k\,y^2}{kY_0}\sqrt{2\pi\sigma^2 t}}\left[\exp\left(-\frac{\left(\frac{1}{ky}-\frac{1}{kY_0}\right)^2}{2\sigma^2 t}\right)-\exp\left(-\frac{\left(\frac{1}{ky}+\frac{1}{kY_0}\right)^2}{2\sigma^2 t}\right)\right]\\
&= \frac{Y_0}{\sigma k\sqrt{2\pi t}\,y^3}\left[\exp\left(-\frac{1}{2\sigma^2k^2t}\left(\frac{1}{y}-\frac{1}{Y_0}\right)^2\right)-\exp\left(-\frac{1}{2\sigma^2k^2t}\left(\frac{1}{y}+\frac{1}{Y_0}\right)^2\right)\right].
\end{aligned} \tag{D15}$$

This provides Eq. (21) and the theoretical distribution used in Fig. 5(b).